\documentclass{jpsj-suppl}
\usepackage{graphics,graphicx}
\usepackage{times}
\usepackage{color}

\usepackage[top=1.25in, bottom=1.25in, left=1in, right=1in]{geometry}

\title{Dimensional Crossover in Quasi-one-dimensional Spin-1 Heisenberg Antiferromagnets}

\author{Keola~Wierschem$^{1,2}$ and Pinaki~Sengupta$^{1}$}
\inst{$^{1}$School of Physical and Mathematical Sciences, Nanyang Technological University, 21 Nanyang Link, Singapore 637371, Singapore\\$^{2}$International Institute for Complex Adaptive Matter, University of California, 1 Shields Avenue, Davis, CA 95616, USA}
\abst{We study the quenching of the Haldane gap in quasi-one-dimensional systems of weakly coupled spin-1 antiferromagnetic Heisenberg chains. The critical interchain coupling $J_c$ required to stabilize long range magnetic order can be accurately determined from large scale quantum Monte Carlo calculations. Several different geometries of coupled chains are studied, illustrating the dependence of $J_c$ on the coordination of chains. For bipartite geometries, ferromagnetically coupled chains yield similar magnitudes for $J_c$.}
\email{keola@ntu.edu.sg}

\kword{quantum magnetism, Haldane conjecture, low dimensional magnetism}

\begin{document}
\maketitle

\section{Introduction}

The well-known Haldane conjecture states that antiferromagnetic Heisenberg spin chains with integer spin possess an excitation gap~\cite{Haldane1983}. This implies that a system of weakly coupled chains at zero temperature will remain in the gapped Haldane phase until the interchain coupling $J$ reaches a critical value $J_c$ sufficient to quench the Haldane gap and establish long range magnetic order. This theoretical picture has been verified through the discovery of a number of Haldane gap materials, such as Ni(C$_2$H$_8$N$_2$)$_2$NO$_2$(ClO$_4$), abbreviated as NENP\cite{Renard1987}.

The ground state of the spin-1 Heisenberg antiferromagnetic chain has been well established. Following Haldane's initial conjecture that integer spin chains have gapped excitations~\cite{Haldane1983}, numeric work was undertaken to confirm this prediction for the spin-1 case. Early on, Botet and Jullien showed evidence for a gap through a finite size scaling analysis of exact results for finite chains~\cite{Botet1983}. The Haldane gap was later calculated to high precision by White and Huse using the density matrix renormalization group~\cite{White1993}.

Following the discovery of the Haldane gap material NENP, Sakai and Takahashi considered the effect of small interchain interactions on a system of antiferromagnetic Heisenberg chains~\cite{Sakai1989}. Through a mean field treatment of exact results for finite chains, a critical coupling $nJ_c\approx0.51$ was found ($n$ being the coordination of chains). Since mean field theory neglects fluctuations, this value represents a lower bound. Later, Koga and Kawakami~\cite{Koga2000} employed a series expansion technique to determine $J_c$ for hypercubic geometries in two and three dimensions. The resulting values for $nJ_c$ were roughly twice as large as those of Sakai and Takahashi. Meanwhile, Kim and Birgeneau~\cite{Kim2000} and Matsumoto {\it et al.}~\cite{Matsumoto2001} both performed quantum Monte Carlo (QMC) calculations for two dimensional geometries in the quasi-one-dimensional limit and arrived at $nJ_c\approx0.08$. This lies above the mean field result, yet below the series expansion value. Thus, it appears that we can use the mean field and series expansion results as lower and upper bounds, respectively. However, to date the more realistic three dimensional geometries have not been considered with recent powerful QMC methods.

In this work, we use a finite size scaling QMC method to accurately determine the critical interchain coupling $J_c$ of the Haldane to N\'{e}el quantum phase transition in three-dimensional systems of spin-1 Heisenberg antiferromagnetic chains in the quasi-one-dimensional limit. By performing this analysis for different chain coordinations $n$, we are able to show that the quantity $nJ_c$ remains roughly constant, as predicted by mean field theory~\cite{Sakai1989}. We also compare the results for ferro- and antiferro-magnetically coupled chains, which turn out to be nearly identical for bipartite lattices.

\section{Model and Methods}

\begin{figure}
\begin{center}
\includegraphics[clip,trim=0cm 13cm 0cm 1cm,width=\linewidth]{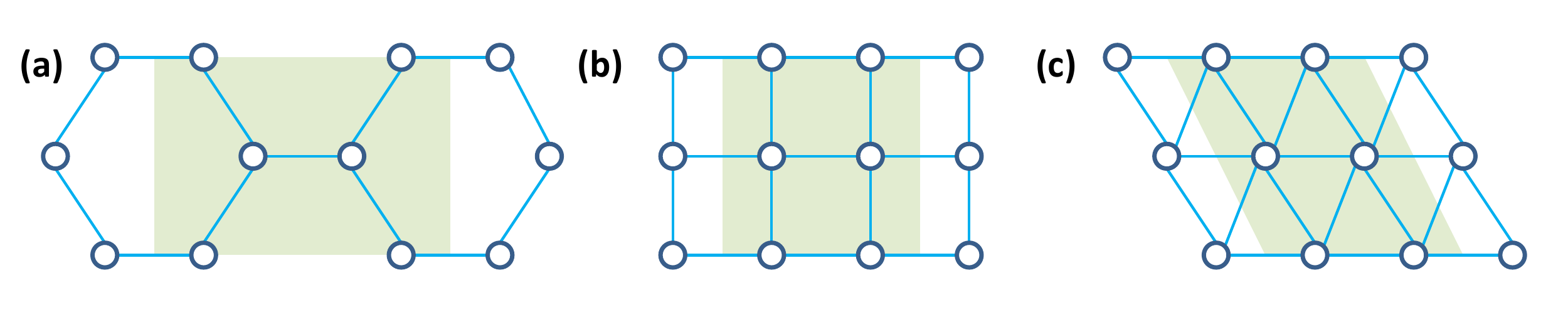}
\end{center}
\caption{(Color online) {\bf (a)} Illustration of the geometric arrangement of chains used in our calculations. Chains are represented by circles that form {\bf (a)} honeycomb ($n=3$), {\bf (b)} square ($n=4$), and {\bf (c)} triangular ($n=6$) lattices. In each case, our non-primitive unit cell is shown as a shaded region.}
\label{geometry}
\end{figure}

We consider a quasi-one-dimensional system of weakly coupled spin-1 Heisenberg antiferromagnetic chains described by the Hamiltonian
\begin{equation}
{\cal H}=J_{\parallel}\sum_{\left<ij\right>_{\parallel}}\vec{S_{i}}\cdot\vec{S_{j}}
		+J_{\perp}\sum_{\left<ij\right>_{\perp}}\vec{S_{i}}\cdot\vec{S_{j}}.
\end{equation}
Here $J_{\parallel}$ is the spin exchange coupling between nearest neighbor spin pairs within a single chain, while $J_{\perp}$ is the spin exchange coupling between nearest neighbor spin pairs on different chains. Without any loss of generality, we set $J_{\parallel}=1$ and use a single parameter $J=J_{\perp}$ to tune the strength of interchain spin coupling. In this work we consider several geometric arrangements of chains with coordination number $3\le n\le6$, as shown in Fig.~\ref{geometry}. Due to the large spatial anisotropy of the spin exchange coupling, we utilize non-cubic simulation cells of dimension $L\times L\times4L$ in order to more rapidly approach the scaling limit~\cite{Sandvik1999}. We find an inverse temperature $\beta=2L$ and system size $N=4L^3=6912$ are sufficient to reach the ground state and thermodynamic limits, respectively.

To investigate the above model, we use the stochastic series expansion QMC method based on a Taylor series expansion of the density matrix in the $S^z$-projected spin basis. Solving the directed loop equations of Sylju\r{a}sen and Sandvik~\cite{Syljuasen2002}, we can minimize bounces in the loop algorithm, leading to efficient global updates. For bipartite geometries (i.e. honeycomb and square), a sublattice rotation can transform the transverse components of the spin exchange interactions from antiferromagnetic to ferromagnetic. This guarantees the Marshall-Peierls sign rule is obeyed, as required to avoid the sign problem in quantum Monte Carlo.

The spin stiffness in $d$ dimensions can be defined in terms of winding numbers~\cite{Pollock1987} by the relation $\rho_s=\frac{3}{2}\frac{\sum_\alpha\langle w_\alpha^2\rangle}{\beta dL^{d-2}}$~\cite{Sandvik1997}. This is a useful observable to distinguish between gapped and gapless states. Since the Haldane state is gapped and the N\'{e}el state is gapless, we can use a finite size scaling method to determine the critical point of the quantum phase transition between these two states. At a critical point in $d$ dimensions, the spin stiffness scales as $\rho_s=L^{2-(d+z)}$, where $z$ is the dynamic critical exponent~\cite{Sandvik1998}. In the present case, $d=3$ and $z=1$, so we expect a quantum phase transition belonging to the four-dimensional Heisenberg universality class. Thus, the crossing point of $\rho_sL^2$ provides an estimate of the critical point for the system under consideration.

\section{Results}

\begin{figure}
\begin{center}
\includegraphics[clip,trim=1cm 4cm 1cm 3cm,width=\linewidth]{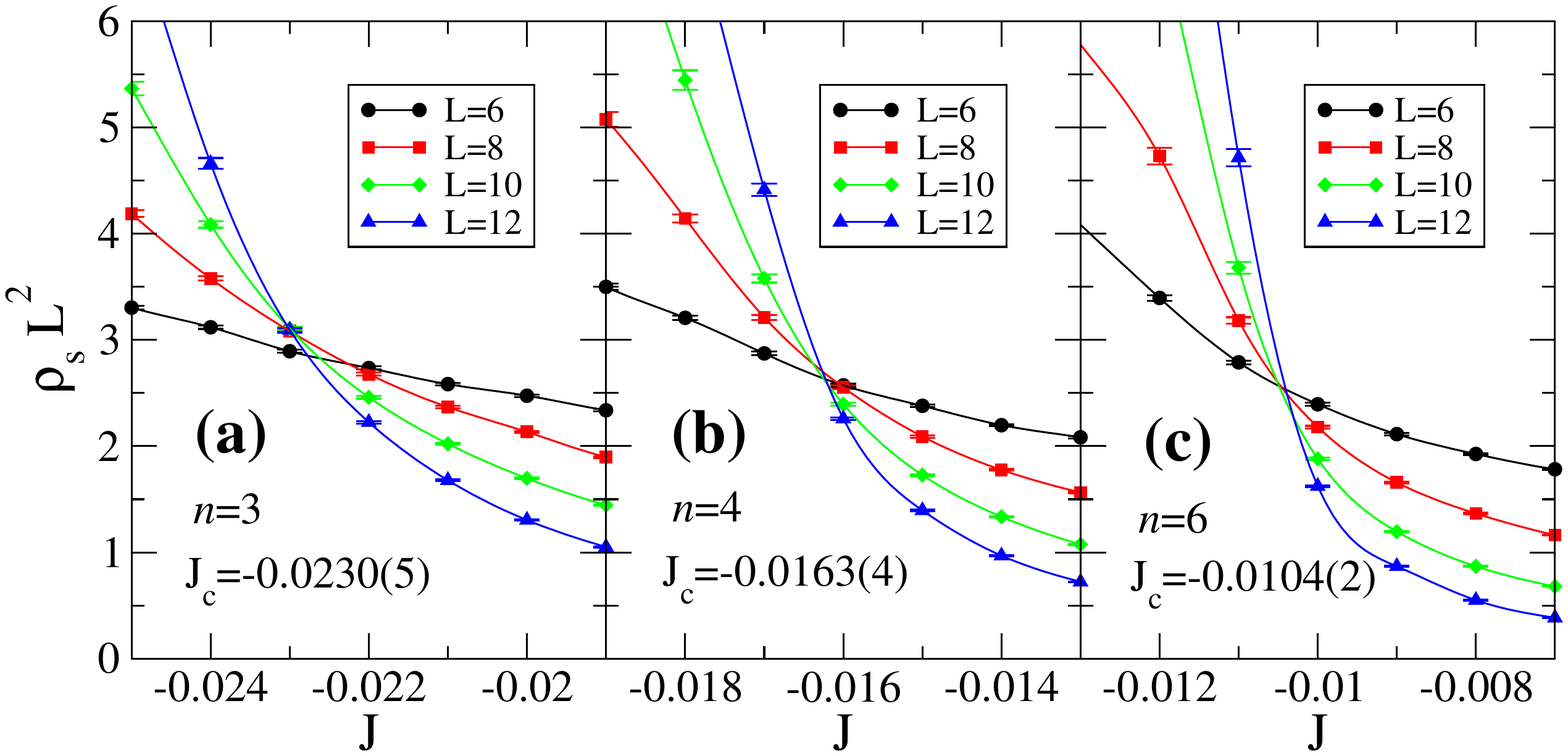}
\end{center}
\caption{(Color online) The scaled spin stiffness $\rho_sL^2$ for ferromagnetically coupled chains in {\bf (a)} honeycomb lattice ($n=3$), {\bf (b)} square lattice ($n=4$) and {\bf (c)} triangular lattice ($n=6$) geometries. The crossing point of $\rho_sL^2$ gives an estimate for the critical coupling $J_c$.}
\label{fig-lattice}
\end{figure}

The effect of geometry on antiferromagnetically coupled chains on bipartite lattices can be determined by QMC calculations. In Fig.~\ref{fig-lattice} we present the results of a finite size scaling analysis of the spin stiffness $\rho_s$ across the Haldane to N\'{e}el phase boundary for honeycomb and square geometries. The crossing point of $\rho_sL^2$ yields values of $J_c=0.0229(6)$ and $J_c=0.0162(4)$, respectively, for these two bipartite geometries. Since the effective dimensionality $d+z$ equals the upper critical dimension, we expect mean field critical exponents for the transition. Using the mean field critical exponent $\nu=1/2$ produces a curve collapse for systems in the critical region, but also indicates the presence of corrections to scaling, which is not unexpected given the spatial anisotropy of our model Hamiltonian.

\begin{figure}
\begin{center}
\includegraphics[clip,trim=1cm 2cm 1cm 4cm,width=\linewidth]{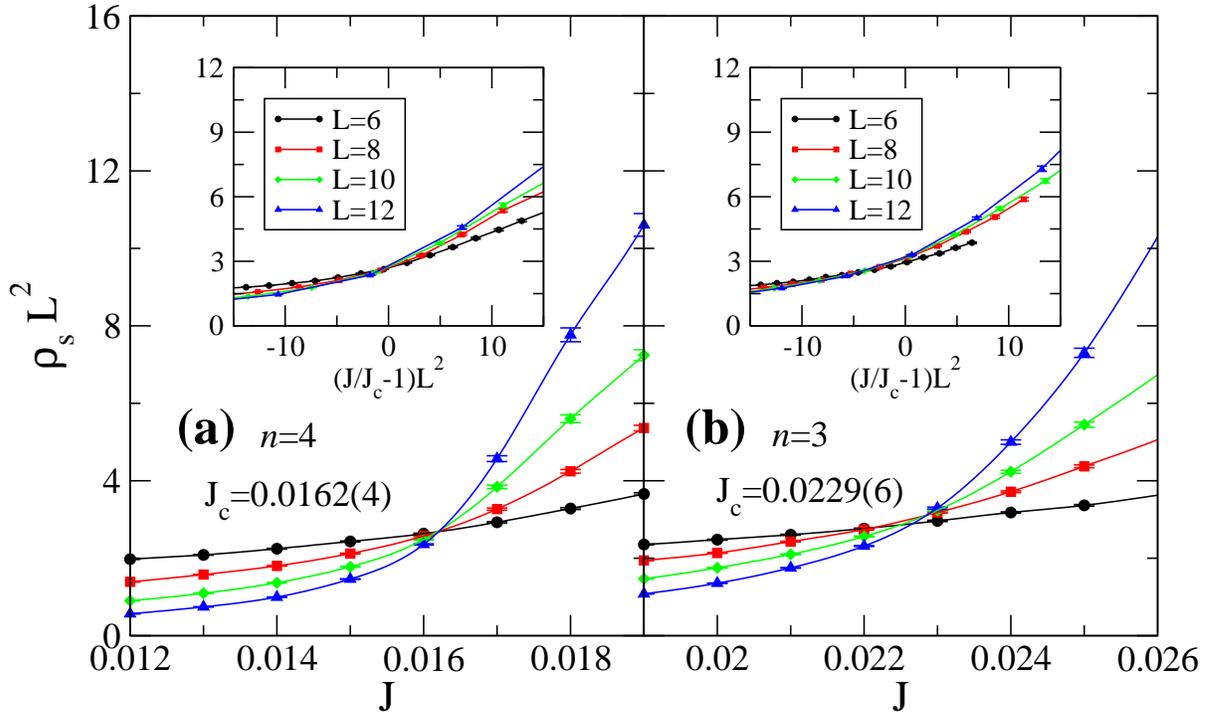}
\end{center}
\caption{(Color online) The scaled spin stiffness $\rho_sL^2$ for antiferromagnetically coupled chains arranged in {\bf (a)} square lattice ($n=4$) and {\bf (b)} honeycomb lattice ($n=3$) geometries. In the main panel, the crossing point of $\rho_sL^2$ gives an estimate for the critical coupling $J_c$. The inset shows finite size scaling curve collapse assuming a mean field value for the critical exponent $\nu$.}
\label{fig-matrix}
\end{figure}

The mean field treatment of Sakai and Takahashi~\cite{Sakai1989} yields a critical coupling that depends only on the coordination number $n$ of the spin chains. Indeed, we find very little variation in $nJ_c$, in qualitative agreement with mean field theory. Comparing our values of $nJ_c$ to past results in Table~\ref{table1}, we find a general agreement. Specifically, our results for $nJ_c$ are larger than the mean field treatment of Sakai and Takahashi~\cite{Sakai1989}, yet smaller than the series expansion of Koga and Kawakami~\cite{Koga2000} or the QMC results of Kim and Birgeneau~\cite{Kim2000} and Matsumoto {\it et al.}~\cite{Matsumoto2001}. This is entirely consistent with the expected role of fluctuations in such systems. Mean field theory neglects fluctuations, which leads to smaller values of $nJ_c$. Additionally, it is known that fluctuations are stronger in lower dimensions, and thus $nJ_c$ will be larger in two dimensional geometries.

\begin{table}
\begin{tabular}{ c c l l l l }
\hline
\hline \\
Source & Method & $d$ & $n$ & $J_{c}$ & $n|J_{c}|$ \\
\hline \\
Sakai and Takahashi~\cite{Sakai1989} & Mean Field & & & & 0.051(1) \\
Koga and Kawakami~\cite{Koga2000} & Series Expansion & 2 & 2 & 0.056(1) & 0.112(2) \\
 & & 3 & 4 & 0.026(1) & 0.104(4) \\
Kim and Birgeneau~\cite{Kim2000} & QMC & 2 & 2 & 0.040(5) & 0.080(10) \\
Matsumoto {\it et al.}~\cite{Matsumoto2001} & QMC & 2 & 2 & 0.043648(8) & 0.087296(16) \\
Present work & QMC & 3 & 3 & 0.0229(6) & 0.0687(18) \\
& & 3 & 4 & 0.0162(4) & 0.0648(16) \\
& & 3 & 3 & -0.0230(5) & 0.0690(15) \\
& & 3 & 4 & -0.0163(4) & 0.0652(16) \\
& & 3 & 6 & -0.0104(2) & 0.0624(12) \\
\hline
\hline
\end{tabular}
\caption{Comparison of critical couplings from divers calculations.}
\label{table1}
\end{table}

The effect of ferromagnetic interchain coupling can be investigated for any geometric arrangement of chains. In Fig.~\ref{fig-lattice} we show results for honeycomb, square, and triangular geometries. As before, the quantity $nJ_c$ varies little between the geometries considered. However, a weak inverse relationship between $nJ_c$ and $n$ is apparent upon closer inspection (see Table~\ref{table1}). Further, we find that the magnitude of the critical coupling is nearly independent of the sign of $J$ on the bipartite lattices. A similar conclusion was also reached for a linear array of coupled chains~\cite{Koga2002}.

\section{Conclusion}

We have performed a QMC study of the quenching of the Haldane gap in quasi-one-dimensional spin-1 Heisenberg antiferromagnets. Using a finite size scaling analysis of the spin stiffness parameter, we determine the critical coupling $J_c$ at which the Haldane gap is quenched and the system transforms into the gapless N\`{e}el state with long range magnetic order. For both ferro- and antiferro-magnetically coupled chains the effect of lattice geometry is shown to be in close qualitative agreement with predictions from mean field theory, with an added weak dependence of $J_c$ on $n$. Finally, the sign of $J$ has little effect for bipartite systems.

\section*{Acknowledgments}
This research used resources of the National Energy Research Scientific Computing Center, which is supported by the Office of Science of the U.S. Department of Energy under Contract No. DE-AC02-05CH11231. One of us, KW, acknowledge the support of the U.S. National Science Foundation I2CAM International Materials Institute Award, Grant DMR-0844115.

\end{document}